\UseRawInputEncoding

\documentclass[runningheads]{llncs}
\usepackage[table]{xcolor}
\definecolor{swotS}{RGB}{226,237,143}
\definecolor{swotW}{RGB}{247,193,139}
\definecolor{swotO}{RGB}{173,208,187}
\definecolor{swotT}{RGB}{192,165,184}
\usepackage[raster]{tcolorbox}
\usepackage[T1]{fontenc}
\usepackage[utf8]{inputenc}
\usepackage{graphicx}
\usepackage{todonotes}

\begin{document}
\title{Interdisciplinary Research with Older Adults in the area of ICT: Selected Ethical Considerations and Challenges}
\author{Kinga Skorupska\inst{1,2,3}\orcidID{0000-0002-9005-0348}
Ewa Makowska \inst{2,1}\orcidID{1111-2222-3333-4444} \and
Anna Jaskulska\inst{1,3}\orcidID{0000-0002-2539-3934} 
}
\authorrunning{Skorupska et al.}
\titlerunning{Selected Ethical Considerations in ICT Research}
%
\institute{Polish-Japanese Academy of Information Technology \and
SWPS University of Social Sciences and Humanities \and Kobo Association
}

\maketitle              
\begin{abstract}
In this paper we analyse, classify and discuss some ethical considerations and challenges related to pursuing exploratory and interdisciplinary research projects in the area of ICT, especially those involving older adults. First, we identify spotlight areas, which are especially prominent in these fields. Next, we explore possible pitfalls interdisciplinary researchers may stumble onto when planning, conducting and presenting exploratory research activities. Finally, some of these are selected and discussed more closely, while related open questions are posed.

\end{abstract}
\section{Introduction}

Ensuring equal access and opportunity, across the board, for people to benefit from and contribute to digital services and content creation is a challenge of an ethical nature in itself. There is still much research to be done in the realm of Information and Communication Technologies (ICT) for older adults with ever more to come as new technologies emerge. Yet, there too are some ethical pitfalls anyone researching this area has to stay aware of. Even the more recent studies in Human-Computer Interaction (HCI) with older adults may be tainted by prevailing stereotypes related to health, social life and ICT skills of participants, as noticed by Vines et al. \cite{articleageoldproblem} and the tendency to consider designing for older adults in terms of mainly accessibility \cite{knowles2020Conflating} resulting in problems in sustainable ICT solutions' development \cite{knowles2018wisdom,knowles2018older}. 

Addressing the widening digital divide \cite{Martinez_2018_DigitalLiteracyOlder} is important not only to allow older adults to engage with cutting-edge technologies but also to empower them to use ICT-based solutions such as online banking, e-health services, e-commerce or e-learning. This is the reason we have researched novel ways of interacting with ICT solutions, such as Smart TV-based interfaces for crowdsourcing \cite{skorupska2018smarttv,skorupska2019smartTV} or ones that are chatbot-based \cite{skorupska2020chatbot}. Furthermore, in the context of Smart Home Technology (SHT) we explored Voice User Interfaces (VUI) \cite{kowalski2019voice,Jaskulska_2021_VisionLoss} and Brain-Computer Interfaces (BCI) \cite{KopecBCI2021}. We have also conducted a VR co-design study with older adults \cite{KOPEC2019277} and created a checklist of factors influencing the Immersive Virtual Reality (IVR) experience for participants of all ages \cite{skorupska2021factors} to make study replication easier. We have also participated in projects evaluating older adults' interaction with online citizen science tasks \cite{zooni2021Interact} and engaged older adults in participatory design activities \cite{kopec2017living}, especially at the frontiers of HCI \cite{KopecPDL2021}.

The considerations we discuss in this paper derive from our research and professional experience, both in the academic and business contexts. It is our hope that they will contribute to the discussion of the adverse effects that rigid scientific practices and insufficient interdisciplinary cooperation may have on the quality of research with older adults at the frontiers of HCI. Researching older adults' preferences regarding digital interaction in an open-minded and ethically sensitive way will help not only mitigate barriers to their use of technology-mediated solutions, but will also help consider older adults' strong suits in new ICT solutions' designs.

\section{Selected Considerations}
There are general ethical considerations related to good research conduct and fair reporting, for example, those outlined in European Code of Conduct for Research Integrity \footnote{ https://allea.org/code-of-conduct/} developed by ALLEA - the European Federation of Academies of Sciences and Humanities, however, each field has its own specificity, and there are clear ethical guidelines on the design of ICT systems \cite{bookethics2015} with Value Sensitive Design (VSD) coming to mind \cite{VSD2007}, just as there are multiple ethical considerations related to research and practice involving participants in general, and older adults in particular. \cite{ethicalfrontiersolderadults2011,ethicsolderadultsfrontirs2016,ethicsolderadults2020}

\subsection{Older Adults and Technology}

However, it is at the intersection of these areas that some challenging ethical questions arise. To bring them to light we have created a SWOT-inspired overview of the selected items in our research field related to its strengths, limitations, opportunities and challenges. The overview is visible in Fig 1. The discussion that follows is based on the interaction of these items and it includes the observations related to challenges and some dilemmas we have encountered in our work so far, as well as ones which feel as though they may become our future concerns.
\\
\begin{tcbraster}[raster columns=2, boxrule=0mm, arc=0mm]
\begin{tcolorbox}[equal height group=A, size=fbox, colback=swotS!60, colframe=swotS!80!black, title=\textsc{strengths}]
\begin{enumerate}
\item creating solutions that foster digital inclusion, serving as a gateway to improve ICT-literacy
\item making it easier to contribute remotely for a group underrepresented online
\item developing solutions with older adults in a participatory manner, drawing insights from them directly
\end{enumerate}

\end{tcolorbox}
\begin{tcolorbox}[equal height group=A, size=fbox, colback=swotW!60, colframe=swotW!80!black, title=\textsc{limitations}]
\begin{enumerate}
\item older adults are a very heterogeneous demographic group
\item easier access to user groups not representative of the general population (interested in ICT)
\item overcoming stereotypes of the researchers
\item overcoming self-stereotypes of the participants
\end{enumerate}

\end{tcolorbox}
\begin{tcolorbox}[equal height group=B, size=fbox, colback=swotO!60, colframe=swotO!80!black, title=\textsc{opportunities}]
\begin{enumerate}
\item exploring the ageing process and the opportunities coming with strong suits of older adults (e.g. increase in crystallized intelligence \cite{mcardle2000modeling})
\item discovering older adults' preferences regarding the interaction with novel technologies
\item designing systems that match older adults' strong suits and aspirations
\item exploring and understanding older adults' diverse needs and motivations
\end{enumerate}

\end{tcolorbox}
\begin{tcolorbox}[equal height group=B, size=fbox, colback=swotT!60, colframe=swotT!80!black, title=\textsc{challenges}]
\begin{enumerate}
\item constant awareness of one's own unconscious biases
\item empowering older adults' to evaluate ICT-solutions with confidence
\item avoiding the reinforcement of the filter bubble
\item using jargon-free clear communication to ensure informed consent
\item designing with very diverse groups of older adults in mind
\item appropriately addressing the user needs that may come with old age (e.g. lower working memory \cite{wolfson2014older})
\item IT system design challenges (user privacy, clear communication, accessibility, bias-free content)
\end{enumerate}

\end{tcolorbox}
\end{tcbraster}

\subsection{Interdisciplinary Cooperation}

Although the ethical path for each researcher may seem clearly paved by high profile publications in each discipline and the best practices taught, at the intersection of multiple disciplines this image begins to crack. Interdisciplinary cooperation is challenging first because of jargon or skills which are meant to be complementary, but often serve to divide into those in-the-know or not (which is just a matter of time and willingness to learn). It is also a challenge to become aware of, bring to light and discuss different internalized research practices, goals and expectations, which may be surprising - but can be the result of formal reporting requirements, best practices considered common for each discipline and skills and individual preferences of the researchers.

\subsubsection{Research priorities}
Such collaboration is also difficult because of the subtle differences in the order of values prioritized, which at first may be difficult to realize. For example, from the point of view of ICT-system design, ensuring that recruitment procedures screen potential users for mild cognitive decline goes against the idea of accessible design and preventing ICT-exclusion. However, it is a common practice in some other disciplines, as has its reasons and its place. But when screening procedures are applied automatically, without taking into account the type of research being done (exploratory) and what is its end goal (preventing digital exclusion) then misunderstandings may appear.

\subsubsection{Vocabulary}
Even the vocabulary may differ. For example, while everyone speaks of "participants", for some fields this term is interchangeable with "users", while in others it is common to think of "subjects" - and this actually has deep implications when it comes to the perception of the reasons for doing research and the attitude towards the participants, which may be objectifying them: if we study users, we observe them to gather insights to design something better for them, if we study subjects, then why are we studying them? It takes us one step further from them, but at what point such distance means detachment rather than objectivity?

\subsubsection{Outdated standards}
Another difficult area is connected to standards and procedures which often feel rigid. This can be felt with standard validated questionnaires, which could use an update. This is both in terms of design and content, as language evolves, habits and expectations change, and what in the past may have indicated a problem, now may be a result of reliance on technology (e.g. not remembering one's location on the map, or the date). At a time when we are armed with best practices used commercially for UX writing, design and accessibility, it is a pity some of these updates are yet to reach the academic practice, and are guarded by the ideas of consistency with past research (should we be consistent with sub-par practices?) or the effort needed to validate new designs. Redesigning study documentation and tools could improve understanding, thus allowing them to better perform their function (by testing a closer approximation of reality, rather than the ability to fill out forms) and address some of the concerns related to informed consent (e.g. using images and diagrams which could improve understanding instead of, or along, blocks of text \cite{altchiLizardKing_2018}).

\subsubsection{Business, Public Institutions and NGOs}
Businesses often view older adults through the lens of stereotypical problems related to health and maintaining a reasonable standard of independent living (AAL -- Ambient and Assisted Living). This happens despite perceiving the potential of the silver economy, as the employees in organizations attempting to address the needs of older adults are often younger and unfamiliar with scientific findings in this area \cite{kopecspiral2018}. This situation is aggravated by diverse expectations and practices \cite{de2019knowledge}. Businesses and industry expect fast iterations and immediate outcomes - reflected in the agile approaches to project management \cite{vidoni2021agile}. Meanwhile, the research process is similar to the waterfall project management methodology \cite{mccormick2012waterfall}. This approach is reinforced by grant applications, which often require researchers to present the whole project plan with expected outcomes and KPIs, the declaration of which, without prior research and verification may be affected by stereotypes. Such focus on KPIs is also conductive to treating project participants, including older adults, in an objectifying way, to engage them just enough to meet the project numbers requirements. A similar problem of focusing on indicators is faced by public institutions and NGOs implementing tasks and projects aimed at older adults. In such a case, there is little space left for an individualized approach to the participants.

\subsection{Participants}

\subsubsection{Information Portioning}
There is a trend in research, encouraged by a strict understanding of a research practice to keep the real point of an experiment hidden \cite{TAI2012218_informedconsent}, to say as little as possible about the research being done. However, this leads to a few problems, even if we disregard the discussion of whether informed consent is possible in such cases. For example, if little information is disclosed participants may attempt to guess what the research may be about, and in this, change their behaviour in an unpredictable way. This happens often with older adults as they expect to be tested on the stereotypical problems, such as balance, hearing, eyesight or cognitive performance. It is even more prominent if task design somehow reinforces this perception. So, if a task somehow involves listening, participants may want to prove to the researcher that their hearing is fine and change the behaviour they exhibit, which may affect the actual point of the experiment.

\subsubsection{Distance and Detachment}
If researchers place themselves too far from the participants either by using objectifying language in participant-facing situations, or really any jargon, or by forgetting to check in with work and life outside of the academia, they may underuse their empathy, as it comes with some cognitive costs \cite{Empathy_2019_ishard}, and loosen their connection to reality. This fault is especially evident with endeavours which are meant to have positive social impact addressing problems in contexts the researchers are not very familiar with. One example here would be designing unrealistic seeming simulations for the experiments \cite{KING198682_unrealisticsimulations} (e.g. based on work problems that do not appear often, using speech examples which are rigid and sound unnatural, creating tasks to evaluate items the general population would have little interest in). Such problems could be addressed by introducing the practice of participatory design, in which members of the target group could co-design research scenarios and evaluate their realness and relevance, to make research closer to life. We believe this co-design step could be a very valuable addition to the experiment design process.

\subsubsection{Managing Expectations}
Yet again, keeping one's defined distance from participants is necessary in some types of research to provide unbiased results, however, when in participatory design the study participants become team members and co-designers, then the appropriate distance is harder to pinpoint. It involves the difficult process of navigating the intricacies of each interaction while always thinking of the good of the participants first.

Staying alert when using unscripted communication is important. Especially if the research project has no direct implications for the lives of the participants engaged with it, it is very important to communicate this clearly, as false expectations of working towards immediate personal benefit may appear, especially among vulnerable groups, such as some older adults. The same consideration was also prominent in our previous project concerning migration \footnote{"Advanced Learning and Inclusive Environment (ALIEN) for Higher Education through greater knowledge and understanding of migration flows in Europe" was a project we conducted between 2016-2019. See more at: https://alienproject.pja.edu.pl/} as it is crucial to not engage the resources of people, especially those with fewer resources, in an unclear context. 

\subsection{Researchers}

\subsubsection{Answering Real Needs}
Just like it is important to verify study design in participatory design workshops, especially when threat of stereotyping is present, there comes one more step, which should actually be taken before that. It is necessary to confront the idea of the research and expected solution with the real needs of the potential study participants, unless conducting basic or foundation research. In the business world this is done by doing marketing research, customer segmentation, interviews, ethnographic studies - which all come before the commitment to projects \cite{hewitt2019does,de2019knowledge}. However, in research, the decision to do a project may be facilitated by available grants and quite often, relies on previous literature analysis, which may be not as applicable to the specific situation where the project will take place. 

\subsubsection{Unconscious Biases}
Another challenge is realizing and facing one's own unconscious biases and entering system or experiment design with a mind free of untested or poorly-backed assumptions. One example we encountered was the expectation that older adults will not perform a longer text transcription as part of a crowdsourcing study - which was shown to be false in the study, only because in the end the task was included \cite{zooni2021Interact}. Actually, it is exactly the constant awareness of our haste to make use of shortcuts and prevailing narratives that can help us spot such study design faults at an early stage, thus allowing us to gain unexpected insights.

\subsubsection{Prioritizing Research Outcomes}
This brings us to the replication crisis and the danger of flawed studies influencing the design choices of future research, thus creating a self-reinforcing cycle of confirmation bias. Again, it is our job as researchers to evaluate the validity of previous studies. This is why there is great value in well performed meta-analyses, for example using the PRISMA approach\footnote{http://www.prisma-statement.org/}. For this reason, the constant need and pressure to do original research is also concerning, as sometimes our resources are best spent on doing a proper meta-analysis to update our understanding allowing us (and others) to design better studies and refresh our curriculum, and necessarily teaching methods, to benefit our students (who ought to be thought as our future colleagues) - which could have the biggest impact on our research area and its progress.

\section{Conclusions}

An important ethical question each scientist ought to answer is their role in the society and, in a broader sense, in the world. To what extent should scientists study what is applicable and practical? How far can researchers follow their passion and interests and at what point these become just an exercise of a curious mind, rather than a way to contribute to social good and positive change? Ideally, these goals would be aligned, and curiosity would lead researchers to socially valuable contributions which are immediately applicable in collaboration with policy makers, businesses and industry.

The barriers we encounter in driving research-informed progress go deep. They are related to the structure of the educational system, research world, incentive design and the broader reality we function in as these tend to be rigid. These are challenges that are often too great for any researcher to single-handedly address, so they are also one of the reasons to maintain a healthy network of collaborators --- both in the academia, in our discipline, in other disciplines and out in the real world, working, aspiring and facing problems. Such people not only can help make our research more relevant, to help us meet today's complex and interdisciplinary challenges. They can also guide us through the intertwined web of ethical considerations to take into account, not only when designing and conducting research, but also taking it beyond the confines of the academia. 

\section*{Acknowledgments}

We would like to thank the many people and institutions gathered together by the distributed Living Lab Kobo and HASE Research Group (Human Aspects in Science and Engineering) for their support of this research. In particular, the authors would like to thank the members of XR Lab Polish-Japanese Academy of Information Technology and Emotion-Cognition Lab SWPS University as well as other HASE member institutions. 

\bibliographystyle{splncs04}
\bibliography{bibliography}

\end{document}